\documentclass[epsfig,12pt]{article}
\usepackage{epsfig,amssymb,euscript,bbold}
\newcommand{\be}{\begin{equation}}
\newcommand{\ee}{\end{equation}}
\newcommand{\bea}{\begin{eqnarray}}
\newcommand{\eea}{\end{eqnarray}}
\newcommand{\ba}{\begin{array}}
\newcommand{\p}[1]{(\ref{#1})}
\newcommand{\ea}{\end{array}}

\def\bbox{{\,\lower0.9pt\vbox{\hrule \hbox{\vrule height 0.2 cm
\hskip 0.2 cm \vrule height 0.2 cm}\hrule}\,}}
\newcommand{\dsl}{\pa \kern-0.5em /}

\newcommand{\nn}{\nonumber \\}

\def\ds{\raise.15ex\hbox{/}\kern-.57em\partial}
\def\Ds{\,\raise.15ex\hbox{/}\mkern-13.5mu D}
\font\mybb=msbm10 at 10pt
\def\bb#1{\hbox{\mybb#1}}

\def\bH {\bb{H}}

\begin{document}

\makeatletter
\renewcommand{\theequation}{\thesection.\arabic{equation}}
\@addtoreset{equation}{section}
\makeatother

\baselineskip 18pt


\begin{titlepage}
\vfill
\begin{flushright}
QMW-PH-00-13\\
RU-NHETC-2000-50\\
hep-th/0011190\\
\end{flushright}

\vfill

\begin{center}
\baselineskip=16pt
{\Large\bf Fivebranes Wrapped On} 
\\
{\Large\bf Associative Three-Cycles} 
\vskip 10.mm
{~Bobby S. Acharya$^{\dagger,1}$, ~Jerome P. Gauntlett$^{*,2}$
and ~Nakwoo Kim$^{*,3}$}\\
\vskip 1cm
{\small\it
 $^\dagger$
Department of Physics\\
Rutgers University\\
126 Freylinghuysen Rd\\
NJ 08854-0849, USA}\\
\vspace{6pt}   
\vskip 1 cm
{\small\it
$^*$
Department of Physics\\
Queen Mary, University of London\\
Mile End Rd, London E1 4NS, UK
}\\
\vspace{6pt}
\end{center}
\vfill
\par
\begin{center}
{\bf ABSTRACT}
\end{center}
\begin{quote}
We construct supergravity solutions corresponding to
fivebranes wrapping associative three-cycles of constant
curvature in manifolds of $G_2$-holonomy. The solutions 
preserve 2 supercharges and are first constructed 
in D=7 gauged supergravity and then lifted to D=10,11. 
We show that the low-energy theory of M-fivebranes 
wrapped on a compact hyperbolic three-space is dual to
a superconformal field theory in D=3 by exhibiting a 
flow to an $AdS_4$ region. {}For IIB-fivebranes wrapped 
on a three-sphere we speculate on a connection with 
spontaneous supersymmetry breaking of pure ${\cal N}$=1 
super Yang-Mills theory in D=3. 
\vfill
\vskip 5mm
\hrule width 5.cm
\vskip 5mm
{\small
\noindent $^1$ E-mail: bacharya@physics.rutgers.edu \\
\noindent $^2$ E-mail: j.p.gauntlett@qmw.ac.uk \\
\noindent $^3$ E-mail: n.kim@qmw.ac.uk \\
}
\end{quote}
\end{titlepage}
\setcounter{equation}{0}

\section{Introduction}

When a brane wraps a supersymmetric cycle
one typically finds a  ``twisted'' field
theory realised on the worldvolume of the brane \cite{bvs}.  
One way to see this is to note that the cycle will 
typically not have a covariantly constant spinor and 
hence supersymmetry must be realised in some twisted fashion. 
The transverse fluctuations of the brane are specified by
sections of  the normal bundle of the brane worldvolume and it is 
the structure of this bundle that gives rise to the twisting.

An investigation of the supergravity/string-theory duals of
such theories was presented in \cite{malnun,malnuntwo} (for related
work see \cite{AO,CV,FS1,FS2,FS3,klebtseyt,klebstrass,cvetic}).  
Consider a supersymmetric spacetime of the form $\mathbb{R}^{1,q}\times M$
and a $q$+$p$-brane wrapping a supersymmetric $p$-cycle $\Sigma_p\subset M$. 
After taking an appropriate limit to decouple 
gravity while keeping the volume of $\Sigma_p$ fixed \cite{malnun,mal,imsy},
one obtains a twisted field theory on the world-volume of the brane
$\mathbb{R}^{1,q}\times \Sigma_p$. It was argued in \cite{malnun} 
that in this decoupling limit the field theory is insensitive 
to the global geometry of $M$: its effect is local and simply determines 
the specific twisted field theory.
At energies low compared to the inverse size of
$\Sigma_p$, these theories then reduce to $D$ = $(q+1)$-dimensional 
field theories.
If we have a large number of branes wrapping the cycle we might expect to 
be able to find supergravity duals for these theories.

The cases analysed in \cite{malnun}
correspond to M-fivebranes and D3-branes 
wrapping Riemann-surfaces that are holomorphically embedded
in Calabi-Yau two- or three-folds. These give rise to
four-dimensional field theories with ${\cal N}$ = 2,1 supersymmetry and 
two dimensional field theories with (4,4),(2,2) supersymmetry, respectively.
One interesting feature of this work is that supergravity
solutions were found with $AdS_5$ and $AdS_3$ regions in the IR,
respectively, providing new AdS/CFT examples. In subsequent work 
IIB fivebranes wrapped on two-spheres leading 
to ${\cal N}$ = 1 
supersymmetry in four dimensions were studied \cite{malnuntwo}.
A fascinating aspect of this work is that it seems to provide
a starting point for analysing the large N-limit of pure 
${\cal N}$ = 1 supersymmetric Yang-Mills (SYM) theory in four-dimensions.

It is natural to extend these investigations by trying to
construct supergravity duals
corresponding to branes wrapping higher dimensional 
supersymmetric cycles. The examples we will focus 
on in this paper are M-fivebranes or IIB fivebranes wrapping
associative 3-cycles in seven-dimensional 
manifolds with $G_2$-holonomy. These configurations 
preserve 1/16 of the supersymmetry and 
hence lead to three-dimensional field theories with 
${\cal N}$=1 supersymmetry, after suitably decoupling gravity.
The 3-cycles will be taken to have constant curvature:
we will consider three-spheres, hyperbolic three-space 
and possible quotients of these spaces by freely acting 
discrete subgroups of the corresponding isometry groups. 
Note that such quotients of hyperbolic space can be compact.

{}Following the strategy in \cite{malnun,malnuntwo} 
we construct the ten and eleven dimensional
solutions, by first constructing solutions in minimal 
D=7 gauged supergravity. When the topological mass vanishes, 
corresponding to NS 
fivebranes, we find explicit solutions. The
solutions are singular both in D=7 and in D=10,11.
{}For the case of the three-sphere, the $SU(2)$-gauge fields in D=7
have a meron form and moreover we do not find a supersymmetric instanton.
This is surprising in the sense that one 
has the reverse situation in Yang-Mills 
theory where the instantons are supersymmetric and the merons are
non-supersymmetric. At the end of the paper we comment on the possibility
of the singularities of the solutions being resolved by a 
non-supersymmetric instanton and speculate on the connection
with spontaneous supersymmetry breaking of pure ${\cal N}$=1 SYM in D=3.
When the topological mass is non-vanishing,
corresponding to M-fivebranes, for the case of
hyperbolic spaces we find a flow to an $AdS_4$ region.
This implies that at low-energies the corresponding
twisted field theory on the M-fivebrane flows to a 
superconformal field theory in D=3, at least in the large $N$ limit. 

The plan of the rest of this paper is as follows. We begin with some 
preliminary discussion of the twisted theories arising from the fivebranes
wrapping associative three-cycles. The supergravity solutions
are presented in section 3 and the paper closes with some 
discussion of our results.

\section{General Comments on the Twisted Fivebrane Theories}

Consider the Type IIB NS 5-brane wrapped on an associative
3-cycle $\Sigma$ in a manifold $M$ of $G_2$ holonomy. The 
world-volume of the fivebrane is then $\mathbb{R}^{1,2}\times \Sigma$. 
The non-trivial part of the spin connection on the worldvolume
is a connection
on the spin bundle $S$ of $\Sigma$. This is an
$SU(2)\subset Spin(1,5)$ bundle. 
The normal bundle to the fivebrane in the $G_2$-manifold 
is four dimensional and given by $N=S\otimes V$  where $S$
is the spinor bundle of $\Sigma$ and $V$ is a rank two 
$SU(2)$-bundle \cite{mclean}. 
{}From this information the 
appropriate twisting can be deduced (see \cite{blauthom}):
one identifies the structure group of $S$, $SU(2)_{\Sigma}$,
with one of the $SU(2)$ factors, $SU(2)_L$ say, in the 
$Spin(4) \cong SU(2)_L\times SU(2)_R$ R-symmetry group of the fivebrane.

The spin content of the twisted theory can thus be specified by giving
the transformations under $Spin(2,1)\times SU(2)_D\times SU(2)_R$,
where $SU(2)_D$ is the diagonal of $SU(2)_\Sigma\times SU(2)_L$.
Now recall that the fields of the flat fivebrane consist 
4 scalars transforming as $({\bf 1, 2, 2})$, under 
$Spin(5,1)\times SU(2)_L\times SU(2)_R$,
fermions transforming as $({\bf 4,2,1})$+$({\bf \bar 4,1,2})$ and  
a six-dimensional vector field.
By decomposing the $Spin(5,1)$ representations
into $Spin(2,1)\times SU(2)_\Sigma$ representations we can then deduce
the representations of the twisted theory. We find that
the six-dimensional vector field gives rise to a three-dimensional 
vector field plus three scalars transforming as $({\bf 3,1})$
of $SU(2)_D\times SU(2)_R$. The 4 scalars give rise 
to scalars transforming as $({\bf 2,2})$: they have become sections
of the normal bundle mentioned above. Finally, the fermions 
transform as $({\bf 2,3,1})$ + $({\bf 2,1,1})$ + $({\bf 2,2,2})$ of 
$Spin(2,1)\times SU(2)_D\times SU(2)_R$. The spinors that generate 
the supersymmetries on the NS fivebrane transform in exactly 
the same way and $({\bf 2,1,1})$ are the preserved supersymmetries
corresponding to ${\cal N}$=1 in D=3.

When $N$ IIB NS-fivebranes wrap an associative 3-cycle they 
give rise to this twisted theory with all fields in the 
adjoint of $U(N)$. 
At energies much less than the size of the cycle the theory will reduce
to an ${\cal N}$ = 1 
supersymmetric field theory in D=3. The low-energy degrees of 
freedom correspond to ${\cal N}$ = 1 SYM in D=3 but there
could be extra massless fields arising from zero modes of the normal
bundle: harmonic sections of $S\otimes V$. 
In this paper we are only considering associative 3-cycles that
are 3-spheres, hyperbolic 3-space or quotients of these spaces.
An example of a non-compact $G_2$-holonomy manifold with an
associative $S^3$ was described in \cite{bs,gpp}. The $7$-manifold
is in fact the total space of the spin bundle of $S^3$, $S({S^3})$.
$SU(2)$-bundles on $S^3$ are trivial so $S({S^3})$ is homeomorphic
to $\mathbb{R}^4{\times}S^3$. The associative $S^3$ 
is identified as the zero section of $S$. In this
case, as pointed out in \cite{mclean}, it is obvious that the normal
bundle to this sphere is just $S$ itself 
and hence $V$ must be trivial. One can immediately conclude that 
there are no zero modes since there
are no harmonic spinors on the three-sphere. However, we do not know
if this is the generic situation for associative three-spheres.
Similarly one can consider $S({\mathbb{H}^3})$
which admits a $G_2$-holonomy metric defined on an open subset \cite{bs}. 
Again, $V$ is trivial, but in this case because $\mathbb{H}^3$ 
has negative constant curvature it is possible that harmonic spinors may 
exist. 

When M-fivebranes wrap an associative cycle
there is one dimension which is neither tangent to the 
fivebrane world-volume nor
tangent to the manifold with $G_2$ holonomy. The $R$-symmetry is now
$SO(5)$ but the twisting involves embedding the $SU(2)$ spin
connection in an $SO(4)$ subgroup in the same way as for the
IIB fivebrane. Consequently the analysis above allows us to conclude
that this theory also 
preserves ${\cal N}$=1 supersymmetry in D=3. {}For a single fivebrane
we expect to get a single ${\cal N}$=1 D=3 scalar superfield and possibly some
extra massless states arising from the normal bundle. It is less
clear what we will get when we have $N$ coincident fivebranes since
we do not have an explicit six-dimensional Lagrangian for the fivebrane
theory. If we wrap one of the uncompactified
world-volume directions on a circle we would get the twisted theory 
in 2 spacetime dimensions arising from the $U(N)$ D4-brane theory wrapping
the associative 3-cycle.

\section{Supergravity Solutions}
{}Following \cite{malnun,malnuntwo} our strategy for constructing
D=10 and D=11 supergravity solutions corresponding to 
string theory and M-fivebranes wrapped on associative 3-cycles
of constant curvature is to first construct the 
solutions in ${\cal N}$=1 D=7 gauged supergravity.

The bosonic field content of ${\cal N}$=1 
D=7 gauged supergravity \cite{townvn} 
consists of a metric $g$, dilaton $\phi$, 
a three-form potential $A_3$ and $SU(2)$ 
gauge fields $A\equiv A^a(\tau^a/2)$,
where $\tau^a$ are Pauli-matrices. 
The fermions are made up of a dilatino 
$\lambda$ and a gravitino $\psi_\mu$, each an eight
component complex $SU(2)$ Majorana spinor. 
The Lagrangian
for the bosonic fields in the string-frame is given by
\bea
{\cal L}&=&{\sqrt g}
e^{-2\phi}[(R-{1\over 8}F^a_{\mu\nu}F^{a\mu\nu}+4 \partial\phi^2)
 -({h^2\over 2} e^{-4\phi} -4 h e^{-2\phi} -4)]\nn
&&-{1\over 2}e^{2\phi}*G\wedge G +
{1 \over 4}F^a\wedge F^a\wedge A_3-{h\over 2}G\wedge A_3
\eea
where $G=dA_3$ is the four-form field strength and 
$F=dA+iA\wedge A\equiv F^a (\tau^a/2)$ is
the $SU(2)$ field strength. The Einstein metric is related to the
string metric via $g_E=e^{-4\phi/5} g$. The potential
$ V= e^{4\phi/5}[h^2/2 e^{-4\phi} -4h e^{-2\phi} -4]$ is drawn in Figure 1.
Note that we have set the gauge coupling constant of \cite{townvn} to
$\sqrt 2$ and we have rescaled the topological mass $h$ by a factor of 8
for convenience. When the topological mass vanishes we can dualise the
3-form potential and rewrite the Lagrangian in terms of a 2-form potential
$B$ as
\bea
{\cal L}={\sqrt g}
e^{-2\phi}[R-{1\over 8}F^a_{\mu\nu}F^{a\mu\nu}+4 \partial\phi^2-
{1\over 3}H_{\mu\nu\rho}H^{\mu\nu\rho} +4]
\eea
with $dH ={1\over 8} F^a\wedge F^a$.
Bosonic solutions to the equations of motion preserve supersymmetry
if the supersymmetry variation of the dilatino and gravitino vanish:
\bea\label{killspin}
\delta\lambda&=&[\not \partial \phi -{i\over 4}\Gamma^{\mu\nu} 
{}F_{\mu\nu} +{1\over 48} e^{2\phi} \Gamma^{\mu\nu\rho\sigma}
G_{\mu\nu\rho\sigma} - he^{-2\phi} +1]\epsilon=0\nn
\delta\psi_\mu&=&[D_\mu+i A_\mu -{i\over 2} F_{\mu\rho} 
\Gamma^\rho +{1\over 96} e^{2\phi} 
\Gamma_\mu^{\, \, \, \nu\rho\sigma\delta}
G_{\nu\rho\sigma\delta}-{h\over 4} e^{-2\phi}\Gamma_\mu]\epsilon=0
\eea
where the spinor $\epsilon$ carries an $SU(2)$ index upon
which the Pauli matrices act.

\begin{figure}[!htb]
\vspace{5mm}
\begin{center}
\epsfig{file=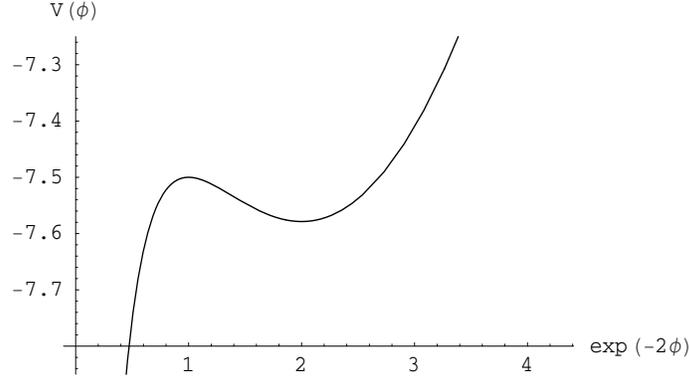,width=9cm,height=5cm}\\
\end{center}
\caption{Scalar potential of D=7 gauged supergravity with $h=1$.}
\label{fig1}
\vspace{5mm}
\end{figure}

To orient ourselves, we first recall some simple configurations
that preserve supersymmetry. {}For vanishing topological mass, 
the linear dilaton solution
\bea
ds^2&=&ds^2(\mathbb{E}^{1,5})+dr^2\nn
\phi&=&-r
\eea
with $F=G=0$ preserves 1/2 of the supersymmetry. Uplifting to D=10
using the formulae in \cite{chamsab,cveticlupope} we get
\bea
ds^2&=&ds^2(\mathbb{E}^{1,5})+dr^2+{1\over 4}(\omega_1^2+\omega_2^2+\omega_3^2)\nn
\phi&=&-r\nn
H^{NS}&=&{1\over 8}\omega_1\wedge\omega_2\wedge\omega_3
\eea
where $\omega_a$ are left invariant one-forms on a 3-sphere which we take to
satisfy $d\omega_1=\omega_2\wedge\omega_3$, and cyclic. 
This solution preserves 1/2 of the ${\cal N}$=1 supersymmetry. 
It is also a supersymmetric solution of IIA/B supergravity
preserving 1/2 supersymmetry and corresponds to the near horizon
limit of the IIA/B NS-fivebrane solution.

When the topological mass is non-vanishing it is more natural to use
the Einstein-frame. The potential for the scalar-field $\phi$
has a supersymmetric maximum at $e^{-2\phi}=1/h$ giving rise to the 
$AdS_7$ solution preserving all supersymmetry:
\bea
ds^2_E&=&R^2[{du^2\over u^2}+u^2 ds^2(\mathbb{E}^{1,5})]\nn
e^{-2\phi}&=&{1\over h}
\eea
with the AdS radius given by $R=2/h^{1/5}$. Using the formulae in 
\cite{lupope} this solution uplifts to $AdS_7\times S^4$ in D=11 
which is the near horizon limit of the M-fivebrane solution.

\subsection{Fivebranes wrapped on three-spheres}

To find more general solutions corresponding to IIB fivebranes and
M-fivebranes wrapped on associative three-spheres, we 
consider an ansatz of the form
\bea
ds^2&=&e^{2 f}[d\xi^2 +d r^2] + a^2 \sigma_1^2 +b^2\sigma_2^2+c^2\sigma_3^2\nn
A&=& \alpha \sigma_1({\tau^1\over 2})+\beta\sigma_2({\tau^2\over 2})+\gamma\sigma_3({\tau^3\over 2})\nn
A_3&=&k \sigma_1\wedge\sigma_2\wedge\sigma_3-l 
d\xi^0\wedge d\xi^1 \wedge d\xi^2
\eea
with $f,\alpha,\beta,\gamma,k,l$ functions of $r$ only and $\sigma_a$ a basis
of left invariant one-forms on $S^3$ satisfying 
$d\sigma_1=\sigma_2\wedge\sigma_3$ and cyclic permutations. Throughout
the paper $d\xi^2$ refers to $ds^2(\mathbb{E}^{1,2})$.
When $a,b,c$ are not all equal, the fivebranes would be wrapping
a squashed three-sphere. Note that in the special case that 
\be\label{conds}
{b^2+c^2-a^2\over 2bc}=\alpha,\qquad
{a^2+c^2-b^2\over 2ac}=\beta,\qquad
{a^2+b^2-c^2\over 2ab}=\gamma
\ee
the $SU(2)$ gauge fields are equal to the components of the
spin connection on the squashed 3-sphere directions.
More precisely, in the frame given by 
$(e^0,\dots,e^6)$ = $(e^f d\xi^0, e^f d\xi^1, e^f d\xi^2, 
e^f dr, a\sigma_1,b\sigma_2,c\sigma_3)$, we then have
${\omega^5}_6=((b^2+c^2-a^2)/2bc)\sigma_1=A^1$ and similarly
${\omega^6}_4=A^2$, ${\omega^4}_5=A^3$. This is the expected 
twisting for associative three-spheres as discussed
in the last section.

Upon substituting this ansatz (without assuming
\p{conds}) into \p{killspin} we have only 
found supersymmetric configurations with non-vanishing 
gauge-fields\footnote{When the topological
mass vanishes there is a supersymmetric solution with vanishing 
gauge-fields, $\alpha$=$\beta$=$\gamma$=0. It has
$f$=$k$=0, $a$=$b$=$c$=constant, $\phi$=$(1+1/4a^2)^{1/2} r$ and
$G$=$(e^{-2\phi}/a )d\xi^0\wedge d\xi^1\wedge d\xi^2\wedge dr$. 
When uplifted to
D=10 it gives rise to the 1/4 supersymmetric IIA/B solution 
corresponding to the near horizon limit of two NS fivebranes 
intersecting on a string that was discussed in \cite{cowdtown}.}
when the four-form is trivial, $k=l=0$,
the squashed 3-spheres are round, $a=b=c$, and $\alpha=\beta=\gamma=1/2$.
Note that these restrictions do indeed satisfy \p{conds}.
Specifically,
\bea
ds^2&=&e^{2f}[d\xi^2 +d r^2] + a^2 (\sigma_1^2 +\sigma_2^2+\sigma_3^2)\nn
A&=& {1\over 2}[\sigma_1({\tau^1\over 2})+\sigma_2({\tau^2\over 2})+\sigma_3({\tau^3\over 2})]\nn
A_3&=&0
\eea
is a supersymmetric solution to the equations of motion provided that
the functions $a,f,\phi$ solve the differential equations:
\bea\label{ode}
{a'\over a}e^{-f}-{1\over 4 a^2} -{h\over 2} e^{-2\phi}&=&0\nn
e^{-f}\phi'-{3\over 16 a^2} -h e^{-2\phi} +1&=&0\nn
2e^{-f} f'-h e^{-2\phi}&=&0
\eea
These configurations preserve 1/8 of the 16 supercharges.
If $\gamma^a$ are gamma matrices with respect to the above mentioned 
frame, the preserved supersymmetries
satisfy 
$\gamma^3\epsilon$=$i\gamma^{56}\tau^1\epsilon$=$i\gamma^{64}\tau^2\epsilon$=$i\gamma^{45}\tau^3$=$\epsilon$ 
(note that the last
condition is implied by the second and third conditions.). 
The spinor $\epsilon$ has a radial dependence given by $\epsilon$=
$e^{f/2}\epsilon_0$ for constant $\epsilon_0$. 
Since the spinors are independent of the coordinates on 
the three-sphere, these solutions are also supersymmetric on $S^3/\Gamma$,
where $\Gamma$ is a discrete subgroup of $SO(4)$, the isometry
group of $S^3$, that acts freely and discontinuously.

It is interesting to note that the gauge-field is half pure-gauge or a meron
\cite{dAFF} (for a recent discussion see \cite{DGI}).
Explicitly, by definition of the left-invariant one-forms $\sigma_a$,
we have $A=-(i/2)\sigma_a(i\tau^a/2)=-(i/2)U^{-1} dU$ where 
$U$ is an arbitrary element of $SU(2)$. In Yang-Mills theory merons
are singular gauge-fields that are not BPS but solve the 
second-order equations of motion. Moreover the 
singularities at the origin and at infinity can be resolved by adding
a half-instanton. Here we have a somewhat reverse situation 
in that the meron is part of a BPS configuration
and we have not been able to find corresponding 
supersymmetric instanton configurations. 
As such it would seem that the 
singularities in the gauge field cannot be resolved by adding 
half-instantons while preserving supersymmetry. 

\subsubsection{Vanishing topological mass}
Let us first consider the case of vanishing topological mass, $h=0$.
In this case we can easily integrate \p{ode} to obtain the 
explicit solution
\bea
ds^2&=&d\xi^2 +d r^2 + {r\over 2} (\sigma_1^2 +\sigma_2^2+\sigma_3^2)\nn
A&=& {1\over 2} [\sigma_1({\tau^1\over 2})+\sigma_2({\tau^2\over 2})
+\sigma_3({\tau^3\over 2})]\nn
e^{2\phi}&=&e^{-2 r} r^{3/4} e^{2\phi_0}\nn
A_3&=&0
\eea
This solution has a curvature singularity at the origin as one
might expect from the singularities of the meron gauge-field.
{}For example, the Ricci-scalar is given by $R=3/r$. 
In  addition to the 3-dimensional Poincare invariance, the
solution is also invariant under SO(4) symmetry. 
(The round 3-sphere is obviously invariant and the gauge fields are up to
an SU(2) gauge transformation.)

This solution can be uplifted to a solution
of ${\cal N}$=1 supergravity in D=10 using the formulae in 
\cite{chamsab,cveticlupope}. Explicitly we get
\bea\label{uplift}
ds^2&=&d\xi^2+dr^2+{r\over 2}(\sigma_1^2+\sigma_2^2+\sigma_3^2)+{1\over 4}
[\nu_1^2+\nu_2^2+\nu_3^2]\nn
e^{2\phi}&=&e^{-2r}r^{3/4}e^{2\phi_0}\nn
H^{NS}&=&{1\over 32}
[\sigma_2\wedge\sigma_3\wedge\nu_1
+\sigma_3\wedge\sigma_1\wedge\nu_2
+\sigma_1\wedge\sigma_2\wedge\nu_3]
+{1\over 8}\nu_1\wedge \nu_2\wedge\nu_3
\eea
with $\nu_a\equiv \omega_a-\sigma_a/2$ and $\omega_a$ 
the left invariant one-forms on a 3-sphere introduced before.
We have directly checked that this solution admits Killing
spinors of ${\cal N}$=1 supergravity in D=10:
\bea
\delta\lambda&=&(\gamma^M\partial_M\phi-{1\over 6}H_{MNP}\gamma^{MNP})
\epsilon=0\nn
\delta\psi_M&=&(D_M-{1\over 4}H_{MNP}\gamma^{NP})\epsilon=0
\eea
provided that $\epsilon$ is constant and satisfies
\be\label{conditions}
\gamma^{3567}\epsilon=\gamma^{3648}\epsilon=\gamma^{3459}\epsilon=\epsilon
\ee
in the frame $(e^0,\dots,e^9)$ = $(d\xi^0,d\xi^1,d\xi^2,dr,
(r/2)^{1/2}[\sigma_1,\sigma_2,\sigma_3],
(1/2)[\nu_1,\nu_2,\nu_3])$. These projections can be recast in the
following elegant way (see, e.g., equations (11) and (78) of \cite{glw}):
\be\label{singlet}
{2\over 3}(\gamma_{ij}+{1\over 4}\psi_{ijkl}\gamma^{kl})\epsilon=0
\ee
where $i,j,k,l\in\{3,4,5,6,7,8,9\}$ and the 
four-form $\psi$ is $G_2$-invariant with non-zero components given by:
\bea
+1&=&\psi_{4578}=\psi_{3459}=\psi_{4679}=\psi_{3567}=\psi_{5689}\nn
-1&=&\psi_{3789}=\psi_{3468}
\eea
Under the decomposition $Spin(9,1)\to Spin(2,1)\times Spin(7)$
the spinors decompose as ${\bf 16}\to ({\bf 2},{\bf 8})$.
We can further decompose $Spin(7)$ under $G_2$ with
${\bf 8}$ $\rightarrow$ ${\bf 1}$+ ${\bf 7}$.
Equation \p{singlet} asserts that $\epsilon$ is $G_2$-invariant.

This solution is also a solution of type IIA and type IIB supergravity,
where it still preserves 2 supercharges, i.e. it now preserves 1/16 
of the supersymmetry. To see this recall that in the string-frame a bosonic
IIA configuration with vanishing Ramond-Ramond fields
is supersymmetric if (see, e.g., \cite{dghw})
\bea
\delta\lambda_\mp&=&\pm(\not\partial\phi\pm{1\over 6}H_{MNP}\gamma^{MNP})
\epsilon_\pm=0\nn
\delta\psi_\pm&=&(D_M\pm{1\over 4}H_{MNP}\gamma^{NP})\epsilon_\pm=0
\eea
We find that the solution breaks all $\epsilon_+$ supersymmetries
and preserves 1/8 of the $\epsilon_-$ supersymmetries. This
is exactly as expected: the $G_2$-holonomy metric will preserve
spinors $\epsilon_\pm$ satisfying \p{conditions}. If we wrap a IIA NS-fivebrane
around the associative 3-cycle in the directions $\{4,5,6\}$ we must impose
$\gamma_{012456}\epsilon_\pm=\epsilon_\pm$ which is only consistent with
$\epsilon_-$.
By explicit calculation, or simply by noting that we can obtain the IIB solution
by performing a trivial T-duality in, the $\xi^1$ or $\xi^2$ direction, 
we conclude that as a solution of the IIB theory it 
also preserves 1/16 of the supersymmetry. We can also
trivially uplift the IIA solution to obtain a 
solution in D=11 preserving 1/16 of the supersymmetry. 

The symmetries of the D=10 solution consist of the D=2+1 Poincare 
invariance as well as $SU(2)^3$ symmetry:
the two left actions for which $\omega$ and $\sigma$ are 
left-invariant and an $SU(2)$ right-action which is the sum
of the two-right actions. These isometries arise from the fact
that the associative 3-cycle is a round three-sphere and that the
normal bundle is not generic.

The asymptotic behaviour of the fivebrane is as one expects for a 
IIB fivebrane wrapping the three-sphere. Presumably the fact that
the three-sphere is getting large is related to the fact that the
D=2+1 gauge coupling has dimension 1/2. 
Note that just as in seven dimensions, the D=10 solution is 
singular as $r\to 0$. {}For example, $H^2$ and 
the Ricci scalar diverge like $1/r^2$. We will return to the 
issue of singularities
in the last section.

Before closing this section, as somewhat of an aside,
we report on a generalisation of 
the solution \p{uplift}. Consider the ansatz
\bea\label{upliftgen}
ds^2&=&d\xi^2+dr^2+a^2(\sigma_1^2+\sigma_2^2+\sigma_3^2)+b^2
[\nu_1^2+\nu_2^2+ \nu_3^2]\nn
H^{NS}&=&
l[\sigma_2\wedge\sigma_3\wedge\nu_1
+\sigma_3\wedge\sigma_1\wedge\nu_2
+\sigma_1\wedge\sigma_2\wedge\nu_3]
+n\nu_1\wedge\nu_2\wedge\nu_3
\eea
This preserves supersymmetry if $l=n/4=C$ for constant $C$, provided
that $a,b,\phi$ satisfy:
\bea\label{odees}
(a^2)'&=&{b\over 2}(1+{8 C \over b^2})\nn
(b^2)'&=&b(1-{b^2\over 4 a^2})(1-{8 C \over b^2})\nn
\phi'&=&{3C \over a^2 b} - {4 C \over b^3}
\eea
{}For $C=1/32$, $b=1/2$ we recover our previous solution \p{uplift}. {}For
$C=0$, after introducing a new radial variable we find
that the non-trivial seven metric is given
\bea\label{geetwo}
ds^2_7=(1-{k^3\over \rho^3})^{-1}d\rho^2 + 
{\rho^2\over 12}(\sigma_1^2+\sigma_2^2+\sigma_3^2)
 +
{\rho^2\over 9} (1-{k^3\over \rho^3})
[\nu_1^2+\nu_2^2+\nu_3^2]
\eea
which is known to be a metric with $G_2$ holonomy \cite{bs,gpp}
and is the one discussed in section 2.
We have not been able to find the exact solution to \p{odees}.
However, we can establish the asymptotic behaviour.
By analysing $db^2/da^2$ we see that for large $a^2$ we have
$b^2\approx 4a^2/3$ and $\phi \approx$ constant.
Using $a$ as a radial variable we then have asymptotically
\be
ds_7^2=12da^2 +a^2(\sigma_1^2+\sigma_2^2+\sigma_3^2)+
{4a^2\over 3} [\nu_1^2+\nu_2^2+\nu_3^2]
\ee
which is the large $\rho$ limit of \p{geetwo}. {}For small 
$a^2$ we have $b^2\approx 1/a$ and
$\phi\approx$ constant$+12 C a$ giving rise to the asymptotic metric
\be
ds_7^2=16a^3da^2 +a^2(\sigma_1^2+\sigma_2^2+\sigma_3^2)+
{1\over a} [\nu_1^2+\nu_2^2+\nu_3^2]
\ee
Note that these solutions are not
solutions of minimal D=7 gauged supergravity because they
have another scalar field active.

\subsubsection{Non-vanishing topological mass}
To find supergravity solutions describing M-theory fivebranes
wrapped on associative three-spheres we need to solve
\p{ode} with $h \ne0$. We have not been able to 
find exact solutions, but it is not too
difficult to establish the asymptotic behaviour of the solutions.
Dividing the second equation by the first and defining
\be
{}F=x^2e^{-2\phi},\qquad x=a^2
\ee
we can obtain the following differential equation 
\be
{dF\over dx}={5F+16 xF\over 4 x+ 8h F}
\ee 
The behaviour of the orbits in the $(F,x)$-plane is illustrated in
figure 2.
It is also useful to note that \p{ode} implies
\be
{df\over dx}={hF\over x^2+2hFx}
\ee

\begin{figure}[!htb]
\vspace{5mm}
\begin{center}
\epsfig{file=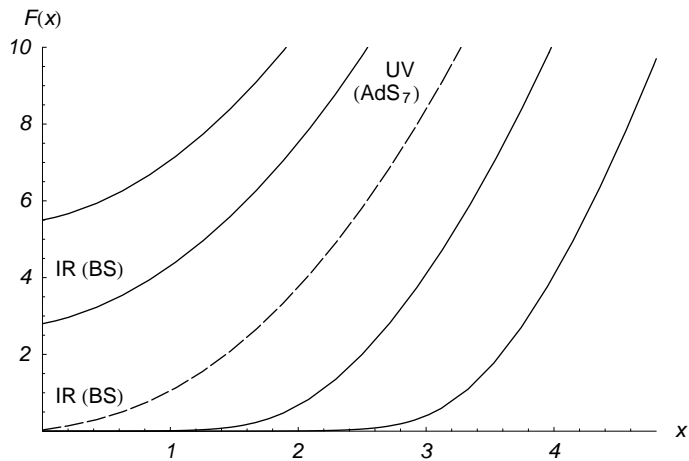,width=9cm,height=6cm}\\
\end{center}
\caption{Behaviour of the orbits for the three-sphere when $h=1$.
The $AdS_7$-type region is at $F\approx x^2$, for large $x$.
The broken line is the separatrix. The singularity in the IR region
is always of the bad type (BS).}
\label{fig2}
\vspace{5mm}
\end{figure}

{}For large $x$ we have $F\approx x^2/h -3x/8h$. Switching the radial
variable from $r$ to $a$ we then deduce that the asymptotic form
of the Einstein metric and scalar are given by
\bea
ds^2_E&=&({2\over h^{1/5}})^2({da^2\over a^2}+
a^2[d\xi^2 +{1\over 4}(\sigma_1^2
+\sigma_2^2+\sigma_3^2)])\nn
e^{-2\phi}&=&{1\over h}-{3\over 8ha^2}
\eea
(we have rescaled the coordinates $\xi$). This is the asymptotic behaviour
that one expects for an M-fivebrane to be wrapped on a three-sphere:
the dilaton is at the supersymmetric maximum of the potential, 
and the metric has the form of $AdS_7$ except that the 
slices of constant $a$ have
$\mathbb{E}^{5,1}$ replaced with $\mathbb{E}^{2,1}\times S^3$.
Moreover, the next to leading order behaviour of the dilaton
is given by $\phi\approx (ln h)/2 +3/(16 a^2)$. 
This corresponds to the insertion of the boundary operator ${\cal O}_\phi$ of 
conformal dimension $\Delta=4$, since the falloff is
like $1/a^{6-\Delta}$. This operator is dual to $\Phi^2$ where
$\Phi$ are the scalars in the tensor multiplet of 
the M-fivebrane theory (see e.g. \cite{dario}). The next
order of the expansion corresponds to the expectation value
of this operator \cite{bkl}.

{}For small $x$, the behaviour of $F$ depends on the value of 
$F(0)$. If it is non-vanishing we have $F\approx F_0+5x/8h$. 
The dilaton and the Einstein metric are then asymptotically
\bea\label{firstbeh}
ds^2_E&=&4 F_0^{2/5}[{a^{22\over 5}\over h^2 F_0^2}da^2
+a^{2\over 5}(d\xi^2 +{1\over 4}(\sigma_1^2
+\sigma_2^2+\sigma_3^2))]\nn
e^{-2\phi}&=&{F_0\over a^4}
\eea
If $F(0)$ vanishes we generically have $F\approx F_0x^{5/4}$ and
\bea
ds^2_E&=&F_0^{2/5}[16a^{7\over 5}da^2
+a^{-{3\over 5}}d\xi^2 +a^{7\over 5}(\sigma_1^2
+\sigma_2^2+\sigma_3^2)]\nn
e^{-2\phi}&=&{F_0\over a^{3/2}}
\eea
The behaviour of the separatrix is $F\approx x/8h$, giving
\bea
ds^2_E&=&{1\over (8h)^{2/5}}[({16\over 5})^2 a^{6\over 5}da^2
+a^{-{2\over 5}}d\xi^2 +a^{6\over 5}(\sigma_1^2
+\sigma_2^2+\sigma_3^2)]\nn
e^{-2\phi}&=&{1\over 8ha^{2}}
\eea
All of these metrics are singular. Note that \p{firstbeh} 
has $g_{00}$ decreasing
as one approaches the singularity, while the others have it diverging.
Before we conclude that the former is thus an example of a ``good''
singularity by the criteria of \cite{malnun} (see also \cite{gubser}), 
we should recall that the definition applied to
the D=11 solution. Using the formulae in \cite{lupope},
we see that the D=11 metric will have the form (e.g., for $h=1$)
\be\label{meteleven}
ds^2=\Delta^{1/3} ds^2_E +X^3\Delta^{1/3}d\theta^2 +{1\over 4}
\Delta^{-2/3}X^{-1}\cos^2\theta(\nu_1^2+\nu_2^2+\nu_3^2)
\ee
where $X=e^{2\phi/5}$
\be
\Delta=X^{-4}\sin^2\theta+X\cos^2\theta
\ee
In all cases the $00$ component of the eleven dimensional metric is divergent
and hence the singularities are ``bad'' by the criteria of \cite{malnun}.

{}Finally, we note that for all solutions
$e^{-2\phi}$ starts from the supersymmetric
maximum at $1/h$, decreases in value before turning and
then running off to infinity.

\subsection{Fivebranes wrapped on hyperbolic space}
Let us now more briefly describe what happens when we replace
the three-sphere with possible quotients of hyperbolic 
three-space, $\mathbb{H}^3/\Gamma$. Here $\Gamma$ is a discrete
subgroup of $SO(3,1)\cong PSL(2,\mathbb{C})$, the isometry group of 
$\mathbb{H}^3/\Gamma$, that acts freely and discontinuously. 
This includes cases when $\mathbb{H}^3/\Gamma$ is compact.

Consider the metric ansatz:
\be
ds^2=e^{2 f}[d\xi^2 +d r^2] + 4{a^2\over z^2}(dz^2+dx^2+dy^2)
\ee
where $(z,x,y)$ are local coordinates on $\mathbb{H}^3$.
We set the four-form $G$ to zero and take the $SU(2)$ gauge fields
$A^a$ to be specified in terms of the spin connection via
\bea
A^1&=&{\omega^5}_6=0\nn
A^2&=&{\omega^6}_4=-{1\over 2a}e^6\nn
A^3&=&{\omega^4}_5={1\over 2a}e^5
\eea
using the frame $(e^f(d\xi^0,d\xi^1,d\xi^2),(2a/z)(z,x,y))$.
These configurations preserve 1/8 of the supersymmetry if
\bea\label{newode}
{a'\over a}e^{-f}-{1\over 4 a^2} +{h\over 2} e^{-2\phi}&=&0\nn
e^{-f}\phi'-{3\over 16 a^2} +h e^{-2\phi} -1&=&0\nn
2e^{-f} f'+h e^{-2\phi}&=&0
\eea
The spinors satisfy $i\gamma^{65}\tau^1\epsilon$=$i\gamma^{46}\tau^2\epsilon$=
$i\gamma^{54}\tau^3$=$\gamma^3\epsilon$=$-\epsilon$ and their 
radial dependence is again given by $\epsilon=e^{f/2}\epsilon_0$ for
constant $\epsilon_0$. 
Since the spinors are independent of the coordinates 
on $\mathbb{H}^3$, these solutions are
also supersymmetric on $\mathbb{H}^3/\Gamma$.

{}For vanishing topological mass, $h=0$, these equations can
be integrated to give metric and dilaton:
\bea
ds^2&=&d\xi^2 +d r^2 + {2r\over z^2}(dz^2+dx^2+dy^2)\nn
e^{2\phi}&=&e^{2 r} r^{-3/4} e^{2\phi_0}
\eea
This solution can then be easily uplifted to D=10. 
Apart from the change in sign of the dilaton this is very similar
to the case of the three-sphere. The solution is singular both
in D=7 and in D=10.

{}For non-vanishing topological mass things are quite different
from the case of the three-sphere. We first note that the
differential equations \p{newode} admit the exact solution
$a^2=5/16$, $e^{-2\phi}=8/5h$ and $e^f=5/4r$. In Einstein-frame 
we have 
\be\label{adsf}
ds^2=\left({8\over 5 h}\right)^{2\over 5}\left[{25\over 16 r^2}
(d\xi^2 +dr^2) +{5\over 4 z^2}(dz^2+dx^2+dy^2)\right]
\ee
corresponding to $AdS_4\times \mathbb{H}^3/\Gamma$. This can easily be lifted
to D=11 using \p{meteleven}, with $\nu_a=\omega_a-A_a$,
and the formula for the four-form in \cite{lupope}.

We can get further insight into the solutions of  \p{newode} by again
analysing the differential equation
for $F=x^2e^{-2\phi}$, $x=a^2$. We now have
\be
{dF\over dx}={-5F+16 xF\over -4 x+ 8h F}
\ee 
Notice that this is the same equation as \p{ode} after
$x\to -x$. The behaviour of the orbits for the region
of interest here is illustrated in figure 3.

\begin{figure}[!htb]
\vspace{5mm}
\begin{center}
\epsfig{file=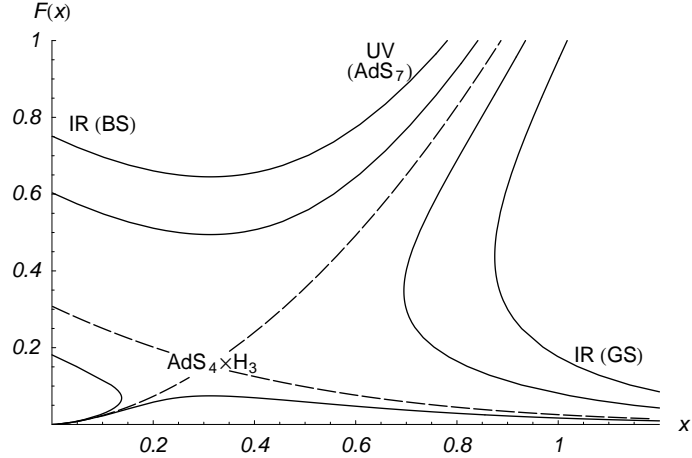,width=9cm,height=6cm}\\
\end{center}
\caption{Behaviour of the orbits for hyperbolic spaces when $h=1$. 
The $AdS_7$-type
region is at $F\approx x^2$, for large $x$, and flows to the IR fixed 
point or the good (GS) and bad (BS) singularities in the IR. 
The broken lines are the separatrices.}

\label{fig3}
\vspace{5mm}
\end{figure}

{}For large $x$ there are solutions that behave like 
$F\approx x^2/h +3x/8h$ which
give rise to the asymptotic solution
\bea\label{asy}
ds^2_E&=&({2\over h^{1/5}})^2({da^2\over a^2}+
a^2[d\xi^2 +{1\over z^2}(dz^2+dx^2+dy^2)])\nn
e^{-2\phi}&=&{1\over h}+{3\over 8ha^2}
\eea
as one expects for an M-fivebrane wrapping the hyperbolic space.
The next to leading order behaviour of the dilaton is now
$\phi\approx (ln h)/2 -3/(16 a^2)$, again corresponding to
the insertion of the boundary operator ${\cal O}_\phi$ of dimension
$\Delta=4$.

There are three different types of behaviour of the these solutions as
$x$ decreases. {}Firstly, there is an orbit that ends up at the
solution \p{adsf}. When $\mathbb{H}^3/\Gamma$ is compact, this orbit
thus corresponds to a flow ``across dimensions''
from the $AdS_7$-type region \p{asy} to an $AdS_4\times \mathbb{H}^3/\Gamma$
region. This implies that the twisted field theory living on
a M-fivebrane wrapped on compact $\mathbb{H}^3/\Gamma$ flows in the far IR
to a new superconformal theory (at least for large $N$) 
whose dual is described by \p{asy}. 

There is also a class of orbits in which for small
$x$, $F$ asymptotes to a constant, $F_0$. These solutions
give rise to the asymptotic metric for small $x$ of the form
\bea
ds^2_E&=&4 F_0^{2/5}[{a^{22\over 5}\over h^2 F_0^2}da^2
+a^{2\over 5}(d\xi^2 +{1\over z^2}(dz^2+dx^2+dy^2)]\nn
e^{-2\phi}&=&{F_0\over a^4}
\eea
These have a similar structure to \p{firstbeh}. In particular,
although $g_{00}$ is decreasing as we approach the singularity
at $a=0$, the $00$ component of the uplifted D=11 metric is 
divergent and hence these are ``bad'' by the criteria of \cite{malnun}.
{}For these flows $e^{-2\phi}$ monotonically increases from $1/h$ to
infinity.

{}Finally there is another class of orbits in which $F$ decreases
as a function of $x$ then turns back on itself and decreases to
zero for large $x$. At the end of these orbits the large $x$
behaviour for $F$ is of the form $F\approx F_0 e^{-4x}$.
Although $a$ is not a good radial coordinate along the whole
of these trajectories, it is good enough to describe
the asymptotic behaviour and we find that they have 
``good'' singularities by the criteria of 
\cite{malnun}. {}For these flows $e^{-2\phi}$ begins increasing from $1/h$
before turning and running back to zero.

\section{Discussion}

We have found supergravity solutions that describe 
fivebranes wrapping associative 3-cycles that are either
$S^3$, $\mathbb{H}^3$ or quotients of these spaces.
{}For the M-fivebrane case we determined 
the general asymptotic behaviour of the solutions to the BPS
equations. In the case of the M-fivebrane wrapping a 
hyperbolic three-space we have shown that there is a flow from
an $AdS_7$ type region to an $AdS_4\times \mathbb{H}^3/\Gamma$
solution. For compact $\mathbb{H}^3/\Gamma$ this implies
that at low-energies the wrapped M-fivebrane theory flows to
a superconformal theory in D=3 at least for large $N$. 
It would be interesting to study this theory in more detail. 
We also found a class of
orbits with ``good'' singularities which presumably correspond 
to switching on a vev for the operator ${\cal O}_\phi$. 
{}For all other orbits, both for the three-sphere
and for hyperbolic spaces, the singularities in the IR are ``bad'' by
the criteria of \cite{malnun}. It will be interesting to see if 
and how they can be resolved.

{}For the IIB NS-fivebrane theory we obtained exact solutions to the
BPS equations and they are all singular in the IR. Let us speculate
on how the singularities might be resolved for the case of the three-sphere.
In section 2 we noted that after taking a suitable 
decoupling limit one expects that the IIB  
NS-fivebrane wrapped on an associative three-sphere 
should give rise to ${\cal N}$=1 supersymmetric Yang-Mills 
theory in the IR, at least
for certain $G_2$ manifolds when the associative three-sphere is rigid. 
Witten has shown \cite{witten} that the Witten 
index vanishes for ${\cal N}$=1 
SYM in D=3 with vanishing Chern-Simons coupling 
and has provided circumstantial evidence that
supersymmetry is actually spontaneously broken. 
If this is indeed the case, it is natural to
suggest that our supergravity solution \p{uplift} describes 
this case and that the singularity can only be removed 
in a non-supersymmetric fashion. 
Recalling that the singularity is related to the
meron gauge fields in D=7 gauged supergravity, it is
plausible that the singularity can be removed by a
supersymmetry breaking instanton. 
It would be interesting to construct such supergravity 
solutions and thereby, hopefully,
demonstrate spontaneous breaking of supersymmetry
in pure ${\cal N}$=1 SYM in D=3. 

We should comment that
Witten has also shown \cite{witten}
that for suitable Chern-Simons couplings the Witten
Index in ${\cal N}$=1 
SYM is non-vanishing and hence supersymmetry is preserved. 
One might hope to be able to find supersymmetric gravity solutions 
describing this situation. Since the IIB NS fivebrane includes 
a coupling of the form $F\wedge F\wedge B_{NS}\sim\omega_{CS}(A)\wedge H$,
where $F$ is the field strength of the gauge fields $A$ living on the
fivebrane, 
one expects that the presence of NS three-form flux $H$ on the
three-sphere would give rise to these theories. However, our original ansatz
did allow for this type of possibility, but we did not 
find such a supersymmetric solution. We have not proven that 
our solutions are the only
ones within our ansatz that preserve supersymmetry but we expect that
a more general ansatz is probably needed to find these solutions, 
if they indeed exist. 

In our approach, following \cite{malnun,malnuntwo}, 
we did not start with a manifold
with $G_2$ holonomy and then construct a solution describing a
fivebrane wrapping an associative 3-cycle. Rather we built the 
solution all at once. This then raises the question about which
$G_2$-holonomy manifolds we are considering in our final solutions.
This does not seem to be a straightforward question to answer
as it is not clear how to ``switch off'' the fivebrane flux. Nevertheless,
it appears that the manifolds are $S(S^3/\Gamma)$ or 
$S(\bH^3/\Gamma)$ of \cite{bs,gpp} 
that we discussed in section 2. The evidence
for this is as follows. Firstly, the decoupling limit that we take
leads to a non-compact manifold: only the local description of the
associative three-cycle is important. Secondly, the structure
of the normal bundle of the associative cycles
in $S(S^3/\Gamma)$ or
$S(\bH^3/\Gamma)$ correspond to our solutions. The generic
normal bundle of an associative three-cycle has structure
group $SO(4)\approx SU(2)_L\times SU(2)_R$ and the twisting 
requires an identification of the $SU(2)$ spin-connection
on the cycle with one of the factors,
$SU(2)_L$, say. Our solutions are constructed in minimal gauged supergravity
which only has $SU(2)_L$ gauge fields and hence we can only construct
solutions corresponding to associative three-cycles with
non-generic normal bundles. The normal bundles to the associative three
cycles in $S(S^3/\Gamma)$ or
$S(\bH^3/\Gamma)$ also have the $SU(2)_R$ bundle trivial.
Finally in section 3.1.1 we derived some
generalised BPS equations that contain these manifolds as well as our solutions
as special cases, for the case of vanishing topological mass.
If we compare (3.20) and (3.12) it is interesting to note that
associative 3-sphere in $S(S^3)$ gets shrunk when we add the 
fivebrane while the other asymptotic 
three-sphere that shrunk to zero size at the zero section
gets blown up to finite size. The latter
is necessary in order for there to be non-zero flux transverse
to the wrapped brane. It is also interesting to note that while
the $G_2$  invariant metric on $S(\bH^3/\Gamma)$ is not complete, 
when we add the M-fivebrane
flux we can get the regular solution 
$AdS_4\times \bH^3/\Gamma$ uplifted to D=11.

\bigskip

{\bf Note Added:}

After this work was completed we became aware of \cite{pernicisezgin}
where they also found the $AdS_4\times \bH^3$ solution of minimal
D=7 gauged supergravity. In addition this paper 
presented an $AdS_4\times \bH^3$
solution for maximal D=7 gauged supergravity. This solution is related
to M-fivebranes wrapping special Lagrangian 3-cycles in Calabi-Yau
three-folds as will be shown elsewhere \cite{gkw}.

\section*{Acknowledgements}
We thank Fay Dowker, Nikita Nekrasov, Paul Townsend, Daniel Waldram and
especially Juan Maldacena for helpful discussions.
JPG thanks the EPSRC for partial support.
JPG and NK are supported in part by PPARC through SPG $\#$613.    



\medskip


\begin{thebibliography}{99}


\bibitem{bvs}
M.~Bershadsky, C.~Vafa and V.~Sadov,
{\it D-Branes and Topological Field Theories},
Nucl.\ Phys.\  {\bf B463} (1996) 420,
hep-th/9511222.

\bibitem{malnun}
J.~Maldacena and C.~Nunez,
{\it Supergravity description of field theories on curved manifolds 
and a no  go theorem},
hep-th/0007018.

\bibitem{malnuntwo}
J.~M.~Maldacena and C.~Nunez,
{\it Towards the large n limit of pure N = 1 super Yang Mills},
hep-th/0008001.


\bibitem{AO}  M. Alishahiha and  Y. Oz,
{\it AdS/CFT and BPS Strings in Four Dimensions},
{\em  Phys.Lett.} {\bf  B465} (1999) 136, hep-th/9907206

\bibitem{CV}  M. Cvetic, H. Lu, C.N. Pope, J.F. Vazquez-Poritz,
{\it AdS in Warped Spacetimes}, Phys.Rev. {\bf D62} (2000) 122003, hep-th/0005246.

\bibitem{FS1} A. Fayyazuddin and D. J. Smith,
{\it Localized intersections of M5-branes and four-dimensional superconformal fiel
d theories},
{\em JHEP} {\bf  9904} (1999) 030,
hep-th/9902210.

\bibitem{FS2} 
A. Fayyazuddin and D. J. Smith,
{\it Warped AdS near-horizon geometry of completely localized intersections of M5-branes},
{\em JHEP} 0010 (2000) 023, hep-th/0006060.

\bibitem{FS3} 
B. Brinne, A. Fayyazuddin, S. Mukhopadhyay, D. J. Smith,
{\it Supergravity M5-branes wrapped on Riemann surfaces and their QFT duals},
hep-th/0009047.    

\bibitem{klebtseyt}
I.~R.~Klebanov and A.~A.~Tseytlin,
{\it Gravity duals of supersymmetric SU(N) x SU(N+M) gauge theories},
Nucl.\ Phys.\  {\bf B578} (2000) 123,
[hep-th/0002159].

\bibitem{klebstrass}
I.~R.~Klebanov and M.~J.~Strassler,
{\it Supergravity and a confining gauge theory: Duality cascades and  
$\chi$SB-resolution of naked singularities}, JHEP {\bf 0008} (2000) 052,
[hep-th/0007191].

\bibitem{cvetic}
M.~Cvetic, H.~Lu and C.~N.~Pope,
{\it Brane resolution through transgression},
hep-th/0011023.

\bibitem{mal}
J.~Maldacena,
{\it The large N limit of superconformal field theories and supergravity},
Adv.\ Theor.\ Math.\ Phys.\  {\bf 2} (1998) 231,
hep-th/9711200.

\bibitem{imsy}
N.~Itzhaki, J.~M.~Maldacena, J.~Sonnenschein and S.~Yankielowicz,
{\it Supergravity and the large N limit of theories with sixteen  supercharges},
Phys.\ Rev.\  {\bf D58} (1998) 046004,
hep-th/9802042.


\bibitem{mclean}
R. McLean, {\it Deformations of Calibrated Sub-Manifolds}, 
Comm. Anal. Geom.  6 (1998) 705-747, available at
http://www.math.duke.edu/preprints/1996.html.

\bibitem{blauthom}
M.~Blau and G.~Thompson,
{\it Aspects of N(T) $\ge$ 2 topological gauge theories and D-branes},
Nucl.\ Phys.\  {\bf B492} (1997) 545,
hep-th/9612143.

\bibitem{townvn}
P.~K.~Townsend and P.~van Nieuwenhuizen,
{\it Gauged Seven-Dimensional Supergravity},
Phys.\ Lett.\  {\bf B125} (1983) 41;
L.~Mezincescu, P.~K.~Townsend and P.~van Nieuwenhuizen,
{\it Stability Of Gauged D = 7 Supergravity And The Definition 
Of Masslessness In (Ads) In Seven-Dimensions},
Phys.\ Lett.\  {\bf B143} (1984) 384.

\bibitem{chamsab}
A.~H.~Chamseddine and W.~A.~Sabra,
{\it D = 7 SU(2) gauged supergravity from D = 10 supergravity},
Phys.\ Lett.\  {\bf B476} (2000) 415
hep-th/9911180.

\bibitem{cveticlupope}
M.~Cvetic, H.~Lu and C.~N.~Pope,
{\it Consistent Kaluza-Klein sphere reductions},
Phys.\ Rev.\  {\bf D62} (2000) 064028
hep-th/0003286.

\bibitem{lupope}
H.~Lu and C.~N.~Pope,
{\it Exact embedding of N = 1, D = 7 gauged supergravity in D = 11},
Phys.\ Lett.\  {\bf B467} (1999) 67
hep-th/9906168.

\bibitem{cowdtown}
P.~M.~Cowdall and P.~K.~Townsend,
{\it Gauged supergravity vacua from intersecting branes},
Phys.\ Lett.\  {\bf B429} (1998) 281,
hep-th/9801165.


\bibitem{dAFF}
V.~de Alfaro, S.~Fubini and G.~Furlan,
{\it A New Classical Solution Of The Yang-Mills Field Equations},
Phys.\ Lett.\  {\bf B65} (1976) 163;
C.~G.~Callan, R.~Dashen and D.~J.~Gross,
{\it A Mechanism For Quark Confinement},
Phys.\ Lett.\  {\bf B66} (1977) 375;
C.~G.~Callan, R.~Dashen and D.~J.~Gross,
{\it Toward a theory of the strong interactions},
Phys.\ Rev.\  {\bf D17} (1978) 2717.

\bibitem{DGI}
N.~Drukker, D.~J.~Gross and N.~Itzhaki,
{\it Sphalerons, merons and unstable branes in AdS},
Phys.\ Rev.\  {\bf D62} (2000) 086007
hep-th/0004131.

\bibitem{glw}
J.~P.~Gauntlett, N.~D.~Lambert and P.~C.~West,
{\it Branes and calibrated geometries},
Commun.\ Math.\ Phys.\  {\bf 202} (1999) 571
hep-th/9803216.

\bibitem{dghw}
A.~Dabholkar, J.~P.~Gauntlett, J.~A.~Harvey and D.~Waldram,
{\it Strings as Solitons $\&$ Black Holes as Strings},
Nucl.\ Phys.\  {\bf B474} (1996) 85
hep-th/9511053.

\bibitem{bs}
R.L. Bryant and S. Salamon, 
{\it On the construction of some complete metrics with exceptional holonomy},
Duke Math. J. {\bf 58} (1989) 829.

\bibitem{gpp}
G.~W.~Gibbons, D.~N.~Page and C.~N.~Pope,
{\it Einstein Metrics On $S^3$, $R^3$ And $R^4$ Bundles},
Commun.\ Math.\ Phys.\  {\bf 127} (1990) 529.

\bibitem{dario}
V.~L.~Campos, G.~Ferretti, H.~Larsson, D.~Martelli and B.~E.~Nilsson,
{\it A study of holographic renormalization group flows in d = 6 and d = 3},
JHEP {\bf 0006} (2000) 023,
hep-th/0003151.

\bibitem{bkl}
V.~Balasubramanian, P.~Kraus and A.~Lawrence,
{\it Bulk vs. boundary dynamics in anti-de Sitter spacetime},
Phys.\ Rev.\ {\bf D 59} (1999) 046003,
hep-th/9805171.

\bibitem{gubser}
S.~S.~Gubser,
{\it Curvature singularities: The good, the bad, and the naked},
hep-th/0002160.

\bibitem{witten}
E.~Witten,
{\it Supersymmetric index of three-dimensional gauge theory},
hep-th/9903005.

\bibitem{pernicisezgin}
M.~Pernici and E.~Sezgin,
{\it Spontaneous Compactification Of Seven-Dimensional Supergravity Theories},
Class.\ Quant.\ Grav.\  {\bf 2} (1985) 673.

\bibitem{gkw}
J.P. Gauntlett, N. Kim and D.J. Waldram, 
{\it M-Fivebranes Wrapped on Supersymmetric Cycles}, hep-th/0012195.

\end{thebibliography}
\end{document}